\documentclass{elsart5}

\usepackage{graphicx}
\usepackage{amssymb}
\usepackage{bm}

\begin{document}

\begin{frontmatter}

\title{Microscopic Aspects of Multipole Properties of Filled Skutterudites}

\author{Takashi Hotta\corauthref{cor1}}
\ead{hotta.takashi@jaea.go.jp}
\corauth[cor1]{}
\address{Advanced Science Research Center,
Japan Atomic Energy Agency, Tokai, Ibaraki 319-1195, Japan}

\received{\today}
\revised{}
\accepted{}

\begin{abstract}
We discuss low-temperature multipole states of Nd-based filled skutterudites
by analyzing a multiorbital Anderson model with the use of
a numerical renormalization group method.
In order to determine the multipole state,
we take a procedure to maximize the multipole susceptibility matrix.
Then, it is found that the dominant multipole state is characterized
by the mixture of 4u magnetic and 5u octupole moments.
The secondary state is specified by 2u octupole.
When we further take into account the coupling between $f$ electrons
in degenerate $\Gamma_{67}^{-}$ ($e_{\rm u}$) orbitals and
dynamical Jahn-Teller phonons with $E_{\rm g}$ symmetry,
quadrupole fluctuations become significant at low temperatures
in the mixed multipole state with 4u magnetic and 5u octupole moments.
Finally, we briefly discuss possible relevance of the present results
to actual Nd-based filled skutterudite compounds.
\end{abstract}




\end{frontmatter}

\section{Introduction}

Recently, magnetism and superconductivity of rare-earth and
actinide compounds have attracted renewed attention
in the research field of condensed matter physics \cite{ASR}.
In particular, filled skutterudites, expressed as RT$_4$X$_{12}$
with rare-earth atom R, transition metal atom T, and pnictogen X,
provide us a platform for systematic research of magnetism and
superconductivity of $f^n$-electron systems
with $n$$\ge$2 \cite{Sato,Aoki1},
where $n$ denotes the number of $f$ electrons.

Since RT$_4$X$_{12}$ crystallizes in the cubic structure
with high symmetry of $T_{\rm h}$ point group \cite{Takegahara},
orbital degeneracy remains in general.
Due to the strong spin-orbit coupling in $f$ electrons,
spin-orbital complex degrees of freedom, i.e., $multipoles$,
become active in filled skutterudites.
For instance, a second-order phase transition at 6.5K
in PrFe$_4$P$_{12}$ \cite{Aoki2} has been considered to be
due to antiferro quadrupole ordering \cite{Iwasa}.
Note that a possibility of antiferro hexadecapole order
has been also suggested in PrFe$_4$P$_{12}$ \cite{Nakanishi2}.
In NdFe$_4$P$_{12}$, a significant role of quadrupole at low temperatures
has been suggested from the measurement of elastic constant \cite{Nakanishi}.
A possibility of octupole ordering in SmRu$_4$P$_{12}$ has been
also pointed out from the elastic constant measurement \cite{Yoshizawa}.
Note that the octupole scenario in SmRu$_4$P$_{12}$
has been supported by muon spin relaxation \cite{Hachitani}
and $^{31}$P NMR experiments \cite{Masaki}.
Quite recently, a possibility of antiferro hexadecapole order
has been proposed to understand metal-insulator transition of PrRu$_4$P$_{12}$
\cite{Takimoto}.

Another characteristic issue of filled skutterudites is $rattling$,
i.e., anharmonic vibrations of rare-earth atom around the off-center
position inside the pnictogen cage.
Effects of rattling on low-temperature $f$-electron states have been
recently discussed actively, in particular, with relevance to
magnetically robust heavy-fermion phenomenon
observed in SmOs$_4$Sb$_{12}$ \cite{Sanada,Miyake}.
Concerning the symmetry of vibrations, a possibility of degenerate
$E_{\rm g}$ mode has been suggested in PrOs$_4$Sb$_{12}$ \cite{Goto}.
Since there exists linear coupling between $f$ electrons in degenerate
$\Gamma_{67}^{-}$ ($e_{\rm u}$) orbitals and vibration mode
with $E_{\rm g}$ symmetry, the present author has pointed out
quasi-Kondo phenomenon due to dynamical Jahn-Teller (JT) phonons
\cite{Hotta1a,Hotta1b}.

In this paper, we focus on the case of $n$=3 as typical example
to study low-temperature multipole properties and
the effect of JT phonons on the multipole state of filled skutterudites.
The multiorbital Anderson model constructed based on a $j$-$j$ coupling
scheme is analyzed by a numerical renormalization group method.
Note that the multipole state is determined by the maximization
of the multipole susceptibility.
It is found that the primary multipole state is characterized by
the mixture of 4u magnetic and 5u octupole moments,
while the secondary state is specified by 2u octupole.
When we further include the coupling between $f$ electrons
and JT phonons, we find that quadrupole fluctuations are
significant at low temperatures
in the 4u-5u coupled multipole state.
Finally, we briefly discuss possible relevance of our results to
actual Nd-based filled skutterudite compounds.

\section{Multiorbital Anderson Model}

The local $f$-electron state is described by \cite{Hotta2a,Hotta2b}
\begin{eqnarray}
  \label{Hloc}
  H_{\rm loc} &=&
  \sum_{m,\sigma,m',\sigma} (B_{m,m'}\delta_{\sigma\sigma'}
  \!+\! \lambda \zeta_{m,\sigma,m',\sigma'})
  f_{m\sigma}^{\dag}f_{m'\sigma'} \nonumber \\
  &+& \sum_{m_1 \sim m_4} \! \sum_{\sigma_1,\sigma_2}
  \! I_{m_1,m_2,m_3,m_4}
  f_{m_1\sigma_1}^{\dag} \! f_{m_2\sigma_2}^{\dag}
  \! f_{m_3\sigma_2} \! f_{m_4\sigma_1},
\end{eqnarray}
where $f_{m\sigma}$ is the annihilation operator for $f$ electrons
with spin $\sigma$ and angular momentum $m$(=$-3$,$\cdots$,3),
$\sigma$=+1 ($-$1) for up (down) spin,
$B_{m,m'}$ is the crystalline electric field (CEF)
potential for angular momentum $\ell$=3,
$\delta_{\sigma\sigma'}$ is the Kronecker's delta,
and $\lambda$ is the spin-orbit coupling.
The matrix element $\zeta_{m,\sigma,m',\sigma'}$ is given by
\begin{equation}
\begin{array}{l}
  \zeta_{m,\pm 1,m,\pm 1}=\pm m/2,\\
  \zeta_{m\pm 1,\mp 1,m, \pm 1}=\sqrt{12-m(m \pm 1)}/2,
\end{array}
\end{equation}
and zero for the other cases.
The Coulomb integral $I_{m_1, m_2, m_3, m_4}$ is expressed
by the combination of four Slater-Condon parameters,
$F^0$, $F^2$, $F^4$, and $F^6$ \cite{Hotta2b}.
In this paper, we set $F^0$=10, $F^2$=5, $F^4$=3, and $F^6$=1
in the unit of eV.
For the $T_{\rm h}$ point group, $B_{m,m'}$ is given by three
CEF parameters, $B_4^0$, $B_6^0$, and $B_6^2$
\cite{Takegahara,Hotta2a}.
In the traditional notations \cite{LLW,Hutchings},
they are expressed as
\begin{equation}
  B_4^0 = Wx/15,~
  B_6^0 = W(1-|x|)/180,~
  B_6^2 = Wy/24.
\end{equation}
where $x$ and $y$ specify the CEF scheme for $T_{\rm h}$ point group,
while $W$ determines an energy scale for the CEF potential.

The local Hamiltonian $H_{\rm loc}$ can provide us exact information
on local $f$-electron states, irrespective of the values of
Coulomb interactions and spin-orbit coupling \cite{Hotta2a}.
However, since $H_{\rm loc}$ includes seven orbitals,
we are immediately faced with difficulties for further study of
many-body phenomena in $f$-electron systems.
Thus, it is natural to consider the effective model which describes
well low-energy states of $H_{\rm loc}$.
For the purpose, we have proposed
to exploit a $j$-$j$ coupling scheme \cite{Hotta3a,Hotta3b}.
We set the spin-orbit coupling term as an unperturbed part,
while the CEF potential and Coulomb interaction terms as perturbations.
Then, we obtain the effective model of $H_{\rm loc}$ as \cite{Hotta3b}
\begin{eqnarray}
  \label{Heff}
  H_{\rm eff} \!=\! \sum_{\mu,\nu} {\tilde B}_{\mu,\nu}
  f_{\mu}^{\dag}f_{\nu} \!+\! \sum_{\mu_1 \sim \mu_4}
  {\tilde I}_{\mu_1, \mu_2, \mu_3, \mu_4}
  f_{\mu_1}^{\dag}f_{\mu_2}^{\dag}f_{\mu_3}f_{\mu_4},
\end{eqnarray}
where $f_{\mu}$ is the annihilation operator for
$f$ electron with angular momentum $\mu$(=$-$5/2,$\cdots$,5/2)
in the $j$=5/2 sextet.
The modified CEF potential is expressed as
\begin{equation}
  {\tilde B}_{\mu,\nu}
  ={\tilde B}^{(0)}_{\mu,\nu}+{\tilde B}^{(1)}_{\mu,\nu},
\end{equation}
where ${\tilde B}^{(0)}_{\mu,\nu}$ denotes the CEF potential for $J$=5/2
and ${\tilde B}^{(1)}_{\mu,\nu}$ is the correction in the order of
$W^2/\lambda$.
The effective interaction in eq.~(\ref{Heff}) is given by
\begin{equation}
  {\tilde I}_{\mu_1, \mu_2, \mu_3, \mu_4}=
  {\tilde I}^{(0)}_{\mu_1, \mu_2, \mu_3, \mu_4}+
  {\tilde I}^{(1)}_{\mu_1, \mu_2, \mu_3, \mu_4},
\end{equation}
where ${\tilde I}^{(0)}_{\mu_1, \mu_2, \mu_3, \mu_4}$ is
expressed by three Racah parameters, $E_0$, $E_1$, and $E_2$,
which are related to the Slater-Condon parameters.
Explicit expressions of ${\tilde I}^{(0)}$ by using $E_0$, $E_1$,
and $E_2$ are shown in Ref.~\cite{Hotta3a}.

On the other hand, ${\tilde I}^{(1)}_{\mu_1, \mu_2, \mu_3, \mu_4}$
is the correction term in the order of $1/\lambda$.
Details on this term have been discussed in Ref.~\cite{Hotta3b}.
Here, three comments are in order.
(i) Effects of $B_6^0$ and $B_6^2$ are included as two-body
potentials in ${\tilde I}^{(1)}$.
(ii) The lowest-order energy of ${\tilde I}^{(1)}$ is
$|W|J_{\rm H}/\lambda$, where $J_{\rm H}$ denotes
the original Hund's rule interaction among $f$ electrons.
(iii) The parameter space in which $H_{\rm eff}$ works is
determined by the conditions for the weak CEF, i.e.,
$|W|/J_{\rm H}$$\ll$1 and $|W|J_{\rm H}/\lambda$$\ll$$E_2$.
Since $E_2$ is the effective Hund's rule interaction
in the $j$-$j$ coupling scheme,
estimated as $E_2$$\sim$$J_{\rm H}/49$ \cite{Hotta3a},
we obtain $|W|/\lambda$$\ll$0.02.
Thus, it is allowed to use $H_{\rm eff}$ even for $\lambda$
in the order of 0.1 eV \cite{Hotta3b}, when $|W|$ is set as
a realistic value in the order of $10^{-4}$ eV
for actual $f$-electron materials.

Now we consider the hybridization between $f$ and conduction electrons.
From the band-structure calculations,
it has been revealed that the main conduction band of
filled skutterudites is $a_{\rm u}$ with xyz symmetry \cite{Harima1},
which is hybridized with $f$ electrons in the $\Gamma_5^-$ state
with $a_{\rm u}$ symmetry.
In order to specify the $f$-electron state,
we introduce ``orbital'' index which distinguishes
three kinds of the Kramers doublets,
two $\Gamma_{67}^-$ and one $\Gamma_5^-$.
Here ``a'' and ``b'' denote the two $\Gamma_{67}^-$'s and ``c''
indicates the $\Gamma_5^-$.

Then, the multiorbital Anderson model is given by
\begin{eqnarray}
  \label{Ham}
  H \!=\! \sum_{\bm{k}\sigma}
  \varepsilon_{\bm{k}} c_{\bm{k}\sigma}^{\dag} c_{\bm{k}\sigma}
  \!+\! \sum_{\bm{k}\sigma}
  (V c_{\bm{k}\sigma}^{\dag}f_{{\rm c}\sigma}+{\rm h.c.})
  \!+\! H_{\rm eff} \!+\! H_{\rm eph},
\end{eqnarray}
where $\varepsilon_{\bm{k}}$ is the dispersion of $a_{\rm u}$
conduction electrons with $\Gamma_5^-$ symmetry,
$f_{\gamma\sigma}$ is the annihilation operator
of $f$ electrons on the impurity site with pseudospin $\sigma$
and orbital $\gamma$,
$c_{\bm{k}\sigma}$ is the annihilation
operator for conduction electrons with momentum $\bm{k}$ and
pseudo-spin $\sigma$,
and $V$ is the hybridization between conduction and $f$ electrons
with $a_{\rm u}$ symmetry.
Throughout this paper, we set $V$=0.05 eV.
Note that the energy unit of $H$ is half of the bandwidth of
the conduction band,
which is considered to be in the order of 1 eV,
since the bandwidth has been typically estimated as 2.7 eV
for PrRu$_4$P$_{12}$ \cite{Harima2}.
Thus, the energy unit of $H$ is taken as eV.
To set the local $f$-electron number as $n$=3,
we adjust the $f$-electron chemical potential.

The last term in eq.~(\ref{Ham}) denotes 
the electron-phonon coupling.
Here, the effect of $E_{\rm g}$ rattling is included as
relative vibration of surrounding atoms.
We remark that localized $\Gamma_{67}^-$ orbitals with
$e_{\rm u}$ symmetry have linear coupling with JT phonons with
$E_{\rm g}$ symmetry, since the symmetric representation of
$e_{\rm u}$$\times$$e_{\rm u}$ includes $E_{\rm g}$.
Then, $H_{\rm eph}$ is given by
\begin{eqnarray}
  \label{Heph}
  H_{\rm eph} &=& g (Q_2 \tau_x + Q_3 \tau_z)
   +(P_2^2+P_3^2)/2 \nonumber \\
  &+&(\omega^2/2)(Q_2^2+Q_3^2) 
   +b(Q_3^3-2Q_2^2Q_3),
\end{eqnarray}
where $g$ is the electron-phonon coupling constant,
$Q_2$ and $Q_3$ are normal coordinates for $(x^2-y^2)$-
and $(3z^2-r^2)$-type JT phonons, respectively,
$P_2$ and $P_3$ are corresponding canonical momenta,
$\tau_{x}$=
$\sum_{\sigma}(f_{{\rm a}\sigma}^{\dag}f_{{\rm b}\sigma}
+f_{{\rm b}\sigma}^{\dag}f_{{\rm a}\sigma})$,
$\tau_{z}$=
$\sum_{\sigma}(f_{{\rm a}\sigma}^{\dag}f_{{\rm a}\sigma}
-f_{{\rm b}\sigma}^{\dag}f_{{\rm b}\sigma})$,
$\omega$ is the frequency of local JT phonons,
and $b$ indicates the cubic anharmonicity.
Note that the reduced mass of JT modes is set as unity.
Here we introduce non-dimensional
electron-phonon coupling constant $\alpha$ and
the anharmonic energy $\beta$ as $\alpha$=$g^2/(2\omega^3)$
and $\beta$=$b/(2\omega)^{3/2}$, respectively.

\section{Multipole Susceptibility}

In order to clarify the magnetic properties at low temperatures,
we usually discuss the magnetic susceptibility, but in more general,
it is necessary to consider the susceptibility of multipole moments
such as dipole, quadrupole, and octupole.
The multipole operator is given in the second-quantized form as
\begin{equation}
   \label{Xop}
   X_{\gamma} = \sum_{\mu,\nu}
   (X_{\gamma})_{\mu\nu} f_{\mu}^{\dag}f_{\nu},
\end{equation}
where $X$ denotes the symbol of multipole with the symmetry of
$\Gamma_{\gamma}$ and $\gamma$ indicates a set of indices
for the irreducible representation.
For $j$=5/2, we can define multipole operators
up to rank 5 in general, but we are primarily interested in
multipole properties from the $\Gamma_{67}^-$ quartet.
Thus, we consider multipole moments up to rank 3
in $O_{\rm h}$ symmetry.

Now we show explicit forms of multipole operators
\cite{Shiina,Hotta4}.
As for dipole moments with $\Gamma_{\rm 4u}$ symmetry,
the operators are expressed as
\begin{equation}
\begin{array}{l}
 J_{{\rm 4u}x}=J_x,~J_{{\rm 4u}y}=J_y,~J_{{\rm 4u}z}=J_z,
\end{array}
\end{equation}
where $J_x$, $J_y$, and $J_{z}$ are three angular momentum operators
for $j$=5/2, respectively.
Concerning quadrupole moments, they are classified into
$\Gamma_{\rm 3g}$ and $\Gamma_{\rm 5g}$.
We express the $\Gamma_{\rm 3g}$ quadrupole operators as
\begin{equation}
\begin{array}{l}
  O_{{\rm 3g}u} = (2J_z^2-J_x^2-J_y^2)/2,~
  O_{{\rm 3g}v} = \sqrt{3}(J_x^2-J_y^2)/2.
\end{array}
\end{equation}
For the $\Gamma_{\rm 5g}$ quadrupole, we have the three operators 
\begin{equation}
  \begin{array}{l}
    O_{{\rm 5g}\xi} = \sqrt{3} \, \overline{J_yJ_z}/2,\\
    O_{{\rm 5g}\eta} = \sqrt{3} \, \overline{J_zJ_x}/2,\\
    O_{{\rm 5g}\zeta} = \sqrt{3} \, \overline{J_xJ_y}/2,
  \end{array}
\end{equation}
where the bar denotes the operation of taking all possible
permutations in terms of cartesian components.

Octupole moments are classified into three types as
$\Gamma_{\rm 2u}$, $\Gamma_{\rm 4u}$, and $\Gamma_{\rm 5u}$.
Among them, $\Gamma_{\rm 2u}$ octupole is written as
\begin{equation}
  \begin{array}{l}
    T_{\rm 2u}=\sqrt{15} \, \overline{J_xJ_yJ_z}/6.
  \end{array}
\end{equation}
For the $\Gamma_{\rm 4u}$ octupole, we express the operators as
\begin{equation}
  \begin{array}{l}
    T_{{\rm 4u}x}=(2J_x^3-\overline{J_xJ_y^2}-\overline{J_xJ_z^2})/2,\\
    T_{{\rm 4u}y}=(2J_y^3-\overline{J_yJ_z^2}-\overline{J_yJ_x^2})/2,\\
    T_{{\rm 4u}z}=(2J_z^3-\overline{J_zJ_x^2}-\overline{J_zJ_y^2})/2,
  \end{array}
\end{equation}
while $\Gamma_{\rm 5u}$ octupole operators are given by
\begin{equation}
\begin{array}{l}
  T_{{\rm 5u}x}=\sqrt{15}(\overline{J_xJ_y^2}-\overline{J_xJ_z^2})/6,\\
  T_{{\rm 5u}y}=\sqrt{15}(\overline{J_yJ_z^2}-\overline{J_yJ_x^2})/6,\\
  T_{{\rm 5u}z}=\sqrt{15}(\overline{J_zJ_x^2}-\overline{J_zJ_y^2})/6.
\end{array}
\end{equation}
Note that we redefine the multipole moments so as to satisfy
the orthonormal condition
Tr$(X_{\gamma}X_{\gamma'})$=$\delta_{\gamma\gamma'}$ \cite{Kubo}.

In principle, the multipole susceptibility can be evaluated
in the linear response theory \cite{Hotta4},
but we should note that the multipole moments
belonging to the same symmetry can be mixed.
In order to determine the coefficient of such a mixed multipole moment,
it is necessary to find the optimized multipole state which
maximizes the susceptibility.
Namely, we define the multipole operator as
\begin{equation}
  M = \sum_{\gamma} p_{\gamma}X_{\gamma},
\end{equation}
where the coefficient $p_{\gamma}$ is determined by the eigenstate
with the maximum eigenvalue of the susceptibility matrix, given by
\begin{eqnarray}
  \label{multisus}
  \chi_{\gamma\gamma'}=
  \frac{1}{Z} \sum_{n,m}
  \frac{e^{-E_n/T}-e^{-E_m/T}}{E_m-E_n}
  \langle n | X_{\gamma} | m \rangle
  \langle m | X_{\gamma'} | n \rangle.
\end{eqnarray}
Here $E_n$ is the eigenenergy for the $n$-th eigenstate $|n\rangle$,
$T$ is a temperature, and $Z$ is the partition function given by
$Z$=$\sum_n e^{-E_n/T}$.

In order to evaluate $\chi_{\gamma\gamma'}$
as well as an entropy $S_{\rm imp}$
and a specific heat $C_{\rm imp}$ of $f$ electrons,
we resort to the numerical renormalization group (NRG) method
\cite{NRG1,NRG2}, in which momentum space is logarithmically
discretized to include efficiently the conduction electrons
near the Fermi energy.
In actual calculations, we introduce a cut-off $\Lambda$ for
the logarithmic discretization of the conduction band.
Due to the limitation of computer resources,
we keep $m$ low-energy states.
In this paper, we set $\Lambda$=5 and $m$=3000.
Note that the temperature $T$ is defined as
$T$=$\Lambda^{-(N-1)/2}$ in the NRG calculation,
where $N$ is the number of the renormalization step.
The phonon basis for each JT mode
is truncated at a finite number $N_{\rm ph}$,
which is set as $N_{\rm ph}$=20 in this paper.

\section{Results}

First let us discuss the CEF states on the basis of $H_{\rm eff}$.
In Figs.~1(a) and 1(b), we show the CEF energy levels for $n$=2 and 3,
respectively.
Here we set $\lambda$=0.1 eV, $W$=$-6$$\times$$10^{-4}$ eV,
and $y$=0.3.
Since the effects of $B_6^0$ and $B_6^2$ are included in $H_{\rm eff}$
as two-body potentials, $H_{\rm eff}$ can reproduce well
the CEF energy levels of the local Hamiltonian $H$
in the realistic intermediate coupling region
with $\lambda/J_{\rm H}$ in the order of 0.1.
As for details, see Ref.~\cite{Hotta3b}.

In order to determine the value of $x$ for Nd-based filled skutterudites,
here we recall that in PrOs$_4$Sb$_{12}$, the ground state is
$\Gamma_1^+$ singlet and the excited state is $\Gamma_4^{+(2)}$
triplet with small excitation energy as large as 10 K
\cite{Gamma1a,Gamma1b,Gamma1c,Gamma1d,Gamma1e}.
Such a situation is well reproduced by choosing $x$=0.4 for $n$=2
in Fig.~1(a).
Now we change rare-earth ion from Pr$^{3+}$ ($n$=2) to Nd$^{3+}$ ($n$=3).
In principle, it is not necessary to modify the CEF parameters
even if rare-earth ion is changed,
since the CEF potential is given by the sum of
electrostatic potentials from ligand anions.
Note, however, that the CEF potentials may be changed
due to the substitution of T and/or X in RT$_4$X$_{12}$.

When we set $x$=0.4 for $n$=3 in Fig.~1(b),
it is observed that the ground state for $n$=3 at $x$=0.4
is $\Gamma_{67}^-$ quartet and the first excited state is
$\Gamma_5^-$ doublet with the excitation energy of 0.02 eV.
Experiments for NdOs$_4$Sb$_{12}$ have suggested $\Gamma_{67}^-$ ground
and $\Gamma_5^-$ excited states
with the excitation energy of 220K \cite{Ho}.
It should be remarked that the theoretical CEF energy levels for $n$=3
agree well with experimental results for NdOs$_4$Sb$_{12}$,
by using the CEF parameters deduced
from the CEF energy levels for PrOs$_4$Sb$_{12}$.
We note that in NdFe$_4$P$_{12}$, both the ground and first excited states
have been found to be $\Gamma_{67}^-$ quartet
with the excitation energy of 222 K \cite{Nakanishi}.
However, in any case, the ground state quartet is well separated from
the first excited state both for NdOs$_4$Sb$_{12}$ and NdFe$_4$P$_{12}$.
In such a situation, it is considered that low-temperature multipole
properties are not sensitive to the first excited state.
Thus, in the following discussion, we fix $x$=0.4.

For the purpose to understand the CEF energy levels of NdFe$_4$P$_{12}$,
it is necessary to consider first those of PrFe$_4$P$_{12}$.
Here we note that the hybridization effect has been considered to play
an important role to understand the difference in the CEF energy states
among Pr-based filled skutterudites \cite{Otsuki,Kuramoto1,Kuramoto2}.
In order to determine the CEF energy levels of NdFe$_4$P$_{12}$,
it is also important to include the effect of hybridization
for the case of $n$=3.
Such calculations can be done, in principle, by using the effective model
$H_{\rm eff}$ on the basis of the $j$-$j$ coupling scheme.
It is one of future problems.

\begin{figure}[t]
\centering
\includegraphics[width=\linewidth]{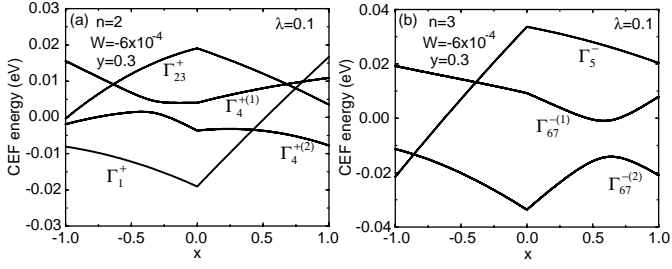}
\caption{CEF energy levels of $H_{\rm eff}$ for (a) $n$=2
and (b) $n$=3.
We set $\lambda$=0.1eV, $W$=$-$6$\times$$10^{-4}$eV,
and $y$=0.3.}
\end{figure}

Now we proceed to the NRG results of $H$.
First let us consider the case without the coupling between
$f$ electrons and JT phonons.
In Fig.~2(a), we show the multipole susceptibility $\chi$.
The dominant multipole moment in the low-temperature region
is the mixture of 4u magnetic and 5u octupole, given by
\begin{equation}
  \label{mom}
  M_{a} = p_a J_{{\rm 4u}a}+q_a T_{{\rm 4u}a} + r_a T_{{\rm 5u}a},
\end{equation}
where we find that $p_a$=$0.989$, $q_a$=$-0.0258$, and $r_a$=$-0.146$
for $a$=$x$, $y$, and $z$.
The mixture of 4u and 5u moments is characteristic of
$T_{\rm h}$ symmetry.
In fact, when we calculate the multipole susceptibility for $O_{\rm h}$
symmetry ($y$=0) using the same parameters except for $y$,
we actually find that $r_a$=0.
It is one of important features of filled skutterudites with
$T_{\rm h}$ symmetry that 4u magnetic moment is
accompanied with 5u octupole.
Note that the magnitude of $r_a$ depends on parameters.
The secondary multipole state is given by 2u octupole.
We also find another mixture of 4u magnetic and 5u octupole moments,
with reduced magnitude of susceptibility.
In Fig.~2(b), we show entropy and specific heat.
At low temperatures, there remains an entropy of $\log 4$,
originating from localized $\Gamma_{67}^-$ quartet,
since we consider the hybridization between $a_{\rm u}$
conduction band and $\Gamma_5^-$ state.
In actuality, there should exist a finite
hybridization between $e_{\rm u}$ conduction bands and
$\Gamma_{67}^-$ states, even if the value is not large compared
with that between $a_{\rm u}$ conduction and $\Gamma_5^-$ electrons.
Thus, the entropy of $\log 4$ should be eventually released.

\begin{figure}[t]
\centering
\includegraphics[width=\linewidth]{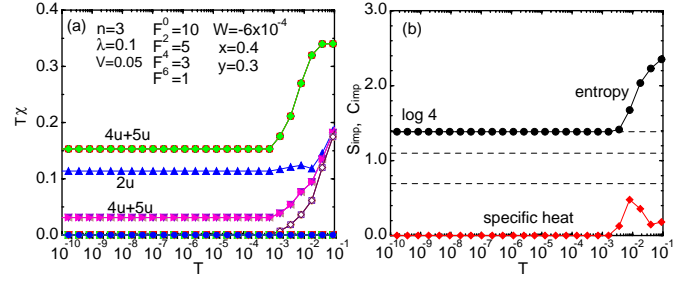}
\caption{(a) $T\chi$ and (b) $S_{\rm imp}$ and $C_{\rm imp}$
vs. temperature for $n$=3 without JT phonons.}
\end{figure}

Next we include the effect of dynamical JT phonons,
but before proceeding to the numerical results,
let us consider intuitively what happens.
In the $j$-$j$ coupling picture, we accommodate three electrons
into the one-electron levels with $\Gamma_5^-$ and $\Gamma_{67}^-$.
Note that $\Gamma_5^-$ is lower than $\Gamma_{67}^-$,
since the ground state of $n$=2 is $\Gamma_1^+$ singlet,
which is mainly composed of doubly occupied $\Gamma_5^-$.
When we accommodate one more electron, it should be put into
$\Gamma_{67}^-$.
Thus, the $\Gamma_{67}^-$ quartet ground state is obtained.
Intuitively, the 4-fold degeneracy is understood by the combination
of spin and orbital degrees of freedom.

Here we consider the JT potential in the adiabatic approximation.
Note that in actuality, the potential is not static,
but it dynamically changes to follow the electron motion.
For $\beta$=0, the potential is continuously degenerate
along the circle of the bottom of the Mexican-hat potential.
Thus, we obtain double degeneracy in the vibronic state
concerning the rotational JT modes
along clockwise and anti-clockwise directions.
When a temperature becomes lower than a characteristic energy
$T^*$, which is related to a time scale to turn
the direction of rotational JT modes,
the entropy of $\log 2$ should be eventually released,
leading to Kondo-like behavior \cite{Hotta1a},
since the specific rotational direction disappears
due to the average over the long enough time.

\begin{figure}[t]
\centering
\includegraphics[width=\linewidth]{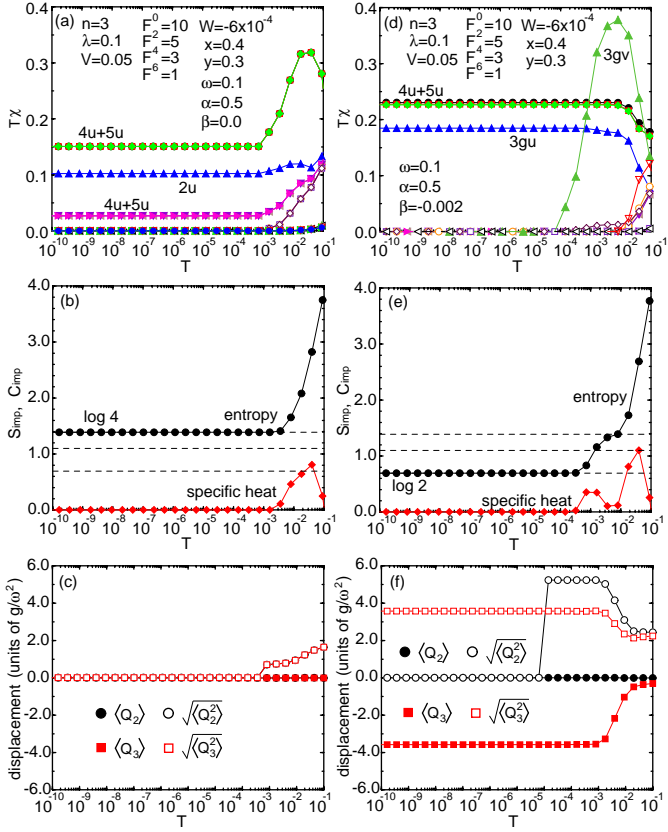}
\caption{(a) $T\chi$, (b) $S_{\rm imp}$ and $C_{\rm imp}$,
and (c) $\langle Q_i \rangle$ and $\sqrt{\langle Q_i^2 \rangle}$
($i$=2 and 3) vs. temperature for $\beta$=0.
(d) $T\chi$, (e) $S_{\rm imp}$ and $C_{\rm imp}$,
and (f) $\langle Q_i \rangle$ and $\sqrt{\langle Q_i^2 \rangle}$
($i$=2 and 3) vs. temperature for $\beta$=$-0.002$.
}
\end{figure}

In Figs.~3(a) and 3(b), we show multipole susceptibilities, entropy,
and specific heat for $\omega$=0.1, $\alpha$=0.5, and $\beta$=0.0.
Note that we suppress the cubic anharmonicity.
We find that the results do not seem to be qualitatively
changed from Figs.~2(a) and 2(b), in spite of the JT active situation.
For the dominant mixed multipole moment,
we find $p_a$=$0.902$, $q_a$=$-0.412$, and $r_a$=$-0.127$
for $a$=$x$, $y$, and $z$ in eq.~(\ref{mom}).
In Fig.~3(c), we show the temperature dependence of average displacements.
At relatively high temperature, we find finite values of
$\sqrt{\langle Q_2^2 \rangle}$ and $\sqrt{\langle Q_3^2 \rangle}$,
while $\langle Q_2 \rangle$=$\langle Q_3 \rangle$=0.
Namely, JT vibrations occur around the origin without displacements.
As the temperature is decreased, such vibrations
are also suppressed and the situation in the low-temperature region
is well described by the electronic model, leading to the similarity
between Figs.~2(a) and 3(a) at low temperatures.

However, when we include the effect of anharmonicity,
three potential minima appear in the bottom of the JT potential
in the adiabatic approximation.
Since the rotational mode should be changed to the quantum tunneling
among three potential minima at low temperatures,
the frequency is effectively reduced in the factor of
$e^{-\delta E/\omega}$,
where $\delta E$ is the potential barrier.
Then, we expect to observe the quasi-Kondo behavior
even in the present temperature range, when we include the
effect of cubic anharmonicity.

In Figs.~3(d) and 3(e), we show multipole susceptibilities, entropy,
and specific heat for $\omega$=0.1, $\alpha$=0.5, and $\beta$=$-0.002$.
For $T$$>$$10^{-4}$, susceptibilities for both
3g quadrupole moments are significant, suggesting that
quadrupole fluctuations are dominant in this temperature region.
However, when the temperature is decreased, $\chi_{{\rm 3g}v}$
is suppressed, while $\chi_{{\rm 3g}u}$ remains at low temperatures.
Instead, the mixed multipole with 4u magnetic and 5u octupole moments
becomes dominant.
Note that $\chi$ for $M_z$ is slightly larger than those for $M_x$ and $M_y$.

Around at $T$=$10^{-3}$, we find a peak in the specific heat,
since an entropy of $\log 2$ is released.
As mentioned above, this is considered to be quasi-Kondo behavior,
originating from the suppression of the rotational mode
of dynamical JT phonons \cite{Hotta1a}.
In this case, the entropy of $\log 2$ concerning orbital degree
of freedom coupled with JT phonons is released,
while there still remains spin degree of freedom
in the localized $\Gamma_{67}^-$ quartet.
In fact, at low temperatures, magnetic susceptibility becomes dominant.

In Fig.~3(f), we show the temperature dependence of average displacements.
For $T$$>$$10^{-5}$, we find $\sqrt{\langle Q_2^2 \rangle}$$\ne$0
and $\sqrt{\langle Q_3^2 \rangle}$$\ne$0,
suggesting that both $Q_2$ and $Q_3$ modes are active.
This is consistent with the finite values of susceptibilities
for $O_{{\rm 3g}u}$ and $O_{{\rm 3g}v}$.
Note that the $Q_3$-type displacement is considered to occur,
since $\langle Q_2 \rangle$=0 and $\langle Q_3 \rangle$$\ne$0.
In the low-temperature region, we find 
$\sqrt{\langle Q_2^2 \rangle}$=$\langle Q_2 \rangle$=0,
while $\sqrt{\langle Q_3^2 \rangle}$=$|\langle Q_3 \rangle|$$\ne$0,
indicating that only $Q_3$-type JT vibration is active
with finite displacement.
This is also consistent with the result that
$\chi_{{\rm 3g}u}$ remains at low temperatures,
since the vibration mode is fixed as $Q_3$-type
after the quasi-Kondo phenomenon occurs.

\section{Discussion and Summary}

In this paper, we have clarified that the magnetic state
with active quadrupole fluctuations appears
in Nd-based filled skutterudites,
when we consider the effect of dynamical JT phonons.
In fact, the existence of degenerate quadrupole moments
has been suggested from the experiment of elastic constant
\cite{Nakanishi}.
In Fig.~3(d), in the temperature region of $T$$>$$10^{-4}$,
we have observed that both $O_{{\rm 3g}u}$ and $O_{{\rm 3g}v}$
become active, although they are not exactly degenerate
due to the effect of JT phonons.
However, the idea of the magnetic state with active quadrupole
fluctuations seems to be consistent with
actual Nd-based filled skutterudites.
Note that in the present NRG calculations, we cannot
conclude the nature of intersite magnetic interaction,
ferromagnetic or antiferromagnetic,
although Nd-based filled skutterudites are ferromagnets.

In Nd-based filled skutterudites such as
NdFe$_4$P$_{12}$ \cite{Sato2} and NdRu$_4$Sb$_{12}$ \cite{Abe},
peculiar behavior of a resistance minimum
in the temperature region higher than a Curie temperature $T_{\rm C}$
has been pointed out.
Quite recently, Np-based filled skutterudite NpFe$_4$P$_{12}$
has been synthesized \cite{DaiAoki}.
Since actinide ion is considered to take a tetravalent state
in the filled skutterudite structure,
NpFe$_4$P$_{12}$ is also classified into the case of $n$=3,
except for the difference between $4f$ and $5f$ electrons states.
In fact, NpFe$_4$P$_{12}$ is also a ferromagnet with $T_{\rm C}$=23 K
and a similar resistance minimum has been observed above $T_{\rm C}$
\cite{DaiAoki}.

In the present paper, we have observed the quasi-Kondo behavior
for the case of $n$=3.
When the temperature is decreased,
an entropy $\log 2$ originating from the double degeneracy
of the vibronic state is released.
In other word, this may be quadrupole Kondo phenomenon,
since quadruple (orbital) degrees of freedom
are tightly coupled with JT phonons,
as understood from Figs.~3(d) and 3(f).
It seems to be premature to conclude the mechanism
only from the present numerical results,
but the quasi-Kondo behavior due to dynamical JT phonons
coupled with orbital (quadrupole) degrees of freedom
may explain qualitatively the resistance minimum phenomenon
in Nd-based filled skutterudites.
Further investigations are required.

As mentioned in the introduction, concerning the mechanism of
magnetically robust heavy-fermion phenomena observed
in SmOs$_4$Sb$_{12}$ \cite{Sanada},
a potential role of phonons has been pointed out from the viewpoint
of the Kondo effect with non-magnetic origin \cite{Miyake}.
In this context, the quasi-Kondo behavior due to the dynamical
JT phonons may be a possible candidate to understand
magnetically robust heavy-fermion phenomenon.
In fact, we have found the quasi-Kondo behavior
also for the case of $n$=5,
but the details of the results on Sm-based filled skutterudites
will be discussed elsewhere \cite{Hotta-Sm}.
Here we emphasize the common feature between
Nd- and Sm-based filled skutterudites with the same $\Gamma_{67}^-$
quartet ground states.
From this viewpoint, it may be interesting to design the
experiment to detect the effect of rattling
in Nd-based filled skutterudites.

In summary, we have discussed the multipole state for $n$=3
by analyzing the multipole Anderson model with the use of
the NRG method.
When we do not consider the coupling between JT phonons and
$f$ electrons in $\Gamma_{67}^-$ quartet, we have found that
the dominant multipole moment is the mixture of 4u magnetic
and 5u octupole.
The secondary multipole state is 2u octupole.
When the coupling with JT phonons is switched and
the cubic anharmonicity is included,
the magnetic ground state includes significant quadrupole
fluctuations and we have found the quasi-Kondo behavior
due to the entropy release concerning the rotational JT mode.

\section*{Acknowledgment}

The author thanks H. Harima, K. Kubo, T. D. Matsuda, Y. Nakanishi,
H. Onishi, and M. Yoshizawa for discussions and comments.
This work has been supported by a Grant-in-Aid for Scientific Research
in Priority Area ``Skutterudites'' under the contract No.~18027016
from the Ministry of Education, Culture, Sports, Science, and
Technology of Japan.
The author has been also supported by a Grant-in-Aid for
Scientific Research (C) under the contract No.~18540361
from Japan Society for the Promotion of Science.
The computation of this work has been done using the facilities
of the Supercomputer Center of Institute for Solid State Physics,
University of Tokyo.



\begin{thebibliography}{00}

\bibitem{ASR}
See, for instance, J. Phys. Soc. Jpn. Suppl. {\bf 75} (2006).

\bibitem{Sato}
H. Sato {\it et al.}:
J. Phys.: Condens. Matter {\bf 15} (2002) S2063.

\bibitem{Aoki1}
Y. Aoki {\it et al.}:
J. Phys. Soc. Jpn. {\bf 74} (2005) 209.

\bibitem{Takegahara}
K. Takegahara, H. Harima and A. Yanase:
J. Phys. Soc. Jpn. {\bf 70} (2001) 1190;
{\it ibid.} {\bf 70} (2001) 3468;
{\it ibid.} {\bf 71} (2002) 372.

\bibitem{Aoki2}
Y. Aoki {\it et al.}:
Phys. Rev. B {\bf 65} (2002) 064446.

\bibitem{Iwasa}
K. Iwasa {\it et al.}:
Physica B {\bf 312-313} (2002) 834.

\bibitem{Nakanishi2}
Y. Nakanishi {\it et al}.: preprint.

\bibitem{Nakanishi}
Y. Nakanishi {\it et al.}:
Phys. Rev. B {\bf 69} (2004) 064409.

\bibitem{Yoshizawa}
M. Yoshizawa {\it et al.}:
J. Phys. Soc. Jpn. {\bf 74} (2005) 2141.

\bibitem{Hachitani}
K. Hachitani {\it et al.}:
Phys. Rev. B {\bf 73} (2006) 052408.

\bibitem{Masaki}
S. Masaki {\it et al.}:
J. Phys. Soc. Jpn. {\bf 75} (2006) 053708.

\bibitem{Takimoto}
T. Takimoto: J. Phys. Soc. Jpn. {\bf 75} (2005) 034714.

\bibitem{Sanada}
S. Sanada {\it et al.}:
J. Phys. Soc. Jpn. {\bf 74} (2005) 246.

\bibitem{Miyake}
S. Yotsuhashi {\it et al.}:
J. Phys. Soc. Jpn. {\bf 74} (2005) 49.

\bibitem{Goto}
T. Goto {\it et al.}:
Phys. Rev. B {\bf 69} (2004) 180511(R).

\bibitem{Hotta1a}
T. Hotta: Phys. Rev. Lett. {\bf 95} (2006) 197201.

\bibitem{Hotta1b}
T. Hotta: Physica B {\bf 378-380} (2006) 51.

\bibitem{Hotta2a}
T. Hotta: J. Phys. Soc. Jpn. {\bf 74} (2005) 1275.

\bibitem{Hotta2b}
T. Hotta: Rep. Prog. Phys. {\bf 69} (2006) 2061.

\bibitem{LLW}
K. R. Lea, M. J. M. Leask and W. P. Wolf:
J. Phys. Chem. Solids {\bf 23} (1962) 1381.

\bibitem{Hutchings}
M. T. Hutchings: Solid State Phys. {\bf 16} (1964) 227.

\bibitem{Hotta3a}
T. Hotta and K. Ueda: Phys. Rev. B {\bf 67} (2003) 104518.

\bibitem{Hotta3b}
T. Hotta and H. Harima: J. Phys. Soc. Jpn. {\bf 75} (2006) No.~12,
in press. See also cond-mat/0602646.

\bibitem{Harima1}
H. Harima and K. Takegahara:
J. Phys.: Condens. Matter {\bf 15} (2003) S2081.

\bibitem{Harima2}
H. Harima {\it et al.}:
J. Phys. Soc. Jpn. Suppl. {\bf 71} (2002) 70.

\bibitem{Shiina}
R. Shiina, H. Shiba and P. Thalmeier:
J. Phys. Soc. Jpn. {\bf 66} (1997) 1741.

\bibitem{Hotta4}
T. Hotta: J. Phys. Soc. Jpn. {\bf 74} (2005) 2425.

\bibitem{Kubo}
K. Kubo and T. Hotta:
J. Phys. Soc. Jpn. {\bf 75} (2006) 013702.

\bibitem{NRG1}
K. G. Wilson: Rev. Mod. Phys. {\bf 47} (1975) 773.

\bibitem{NRG2}
H. R. Krishna-murthy, J. W. Wilkins and K. G. Wilson:
Phys. Rev. B {\bf 21} (1980) 1003.

\bibitem{Gamma1a}
Y. Aoki {\it et al.}:
J. Phys. Soc. Jpn. {\bf 71} (2002) 2098.

\bibitem{Gamma1b}
M. Kohgi {\it et al.}:
J. Phys. Soc. Jpn. {\bf 72} (2003) 1002.

\bibitem{Gamma1c}
T. Tayama {\it et al.}:
J. Phys. Soc. Jpn. {\bf 72} (2003) 1516.

\bibitem{Gamma1d}
K. Kuwahara {\it et al.}:
J. Phys. Soc. Jpn. {\bf 73} (2004) 1438.

\bibitem{Gamma1e}
E. A. Goremychkin {\it et al.}:
Phys. Rev. Lett. {\bf 93} (2004) 157003.

\bibitem{Ho}
P.-C. Ho {\it et al.}:
Phys. Rev. B {\bf 72} (2005) 094410.

\bibitem{Otsuki}
J. Otsuki, H. Kusunose and Y. Kuramoto:
J. Phys. Soc. Jpn. {\bf 74} (2005) 200.

\bibitem{Kuramoto1}
Y. Kuramoto {\it et al.}:
Prog. Theor. Phys. Suppl. {\bf 160} (2005) 134.

\bibitem{Kuramoto2}
Y. Kuramoto {\it et al.}:
J. Phys. Soc. Jpn. Suppl. {\bf 75} (2006) 209.

\bibitem{Sato2}
H. Sato {\it et al.}:
Phys. Rev. B {\bf 62} (2000) 15125.

\bibitem{Abe}
K. Abe {\it et al.}:
J. Phys.: Condens. Matter {\bf 14} (2002) 11757.

\bibitem{DaiAoki}
D. Aoki {\it et al.}:
J. Phys. Soc. Jpn. {\bf 75} (2006) 073703.

\bibitem{Hotta-Sm}
T. Hotta: preprint.

\end{thebibliography}
\end{document}